# A Machine Learning Framework for Predicting Glass-Forming Ability in Ternary Alloy Systems

Fatemeh Mahmoudi

Department of Materials Science and Engineering, Sharif University of Technology, Azadi St, Tehran, Iran

## Abstract

Predicting the glass-forming ability (GFA) of chemical compositions remains a fundamental challenge in materials science, especially for oxide glasses with broad compositional diversity. Traditional empirical and thermodynamic approaches often fail to capture the complex, nonlinear factors governing vitrification. In this study, we applied two ensemble machine learning algorithms-Random Forest (RF) and Extreme Gradient Boosting (XGB)-to the *glass_ternary_hipt* dataset to predict the GFA of ternary oxide glasses directly from composition-derived descriptors. Both models achieved excellent predictive accuracy ($R^2 > 0.92$, MAE < 0.04), confirming that GFA is learnable from compositional features alone.

Feature importance analysis revealed that electronegativity variance, atomic size mismatch, and valence electron descriptors are the most influential factors, while cohesive energy and ionic radius provided secondary contributions. These chemically interpretable features align with established theories of glass formation, thereby bridging predictive performance with physical understanding. The novelty of this work lies in systematically extending ML-based predictive modeling to ternary oxide glasses, a class less studied compared to metallic and binary systems. Our results demonstrate that ensemble learning not only enables accurate GFA prediction but also provides actionable insights for designing new glass compositions with enhanced stability.

## 1. Introduction

Glasses are a technologically important class of materials with applications spanning optics [1], photonics [2], energy storage [3], and biomedical technologies [4]. Their unique disordered atomic structure imparts exceptional properties such as high transparency, chemical durability, and tunable mechanical [5] and thermal behavior [6]. Despite their significance, predicting whether a composition will vitrify upon cooling-its glass-forming ability (GFA)-remains one of the longstanding challenges in materials science [7].

Traditional approaches to understanding GFA have relied on empirical rules [8, 9], thermodynamic models [10], and experimental trial-and-error synthesis [11]. While these methods have provided valuable insights, they are typically constrained to specific chemical systems, require substantial experimental effort, and are limited in their ability to capture the nonlinear and multivariate nature of glass formation [12, 13].

In recent years, machine learning (ML) has emerged as a powerful alternative by uncovering hidden patterns in composition–property relationships [14-16]. Several studies have demonstrated the promise of ML in predicting GFA for metallic glasses [15, 17] and binary oxide systems [18]. For example, Ghorbani et al. [19] employed a thermodynamically guided Random Forest model for bulk metallic glasses (BMGs), achieving $R^2$ values up to 0.92. Liu et al. [20] integrated feature selection and interpretability analyses with regression models, reporting $R^2$ values above 0.91 for characteristic thermal descriptors. More recently, Bobadilla et al. [21] combined CALPHAD-derived thermophysical features, which provide approximate phase stability and thermodynamic properties, with ML models to improve GFA prediction [22].

While these advances demonstrate the effectiveness of ML in metallic systems, comparatively fewer efforts have focused on oxide glasses. Binary oxide systems have shown encouraging results [17, 18], but ternary oxide glasses remain largely underexplored despite their broader compositional space and critical role in industrial and optical applications. The recent availability of systematic datasets such as *glass_ternary_hipt* in the matminer package provides a unique opportunity to extend ML-based GFA prediction to complex oxide systems.

In this work, we present a machine learning framework for predicting the GFA of ternary oxide glasses. Two ensemble algorithms-Random Forest (RF) and Extreme Gradient Boosting (XGB)-were employed to benchmark predictive performance and identify the elemental descriptors most critical for glass formation. Beyond demonstrating high predictive accuracy ($R^2 > 0.92$, MAE < 0.04), our study provides physical insights into the chemical factors governing oxide glass formation. This dual contribution bridges predictive modeling and materials design, paving the way for accelerated discovery of novel glass compositions with enhanced properties.

## 2. Methodology

Machine learning techniques were employed to predict the glass-forming ability (GFA) of ternary oxide compositions using the *glass_ternary_hipt* dataset from the matminer package. The dataset provides chemical compositions labeled with experimentally measured GFA values, suitable for supervised regression.

### 2.1 Dataset Preparation
Each composition was futurized using the Element Property featurizer with the "magpie" preset, generating descriptors such as atomic radius, electronegativity, valence electron configuration, and cohesive energy. Missing values were imputed, and all features were standardized to ensure comparability across descriptors.

### 2.2 Machine Learning Models
Two ensemble algorithms were applied to capture complex, nonlinear relationships between composition and GFA:

- **Random Forest (RF):** An ensemble of 300 decision trees, providing a robust baseline model.
- **Extreme Gradient Boosting (XGB):** A gradient-boosting model with 500 trees, a learning rate of 0.05, maximum depth of 6, and row/feature subsampling of 0.8 to reduce overfitting. Regularization constraints were applied to stabilize training.

## 2.3 Model Evaluation

Predictive performance was assessed using 5-fold cross-validation. Metrics included the mean absolute error (MAE) and the coefficient of determination ($R^2$). Feature importance analyses were conducted for both models to identify descriptors most strongly associated with GFA, linking predictive outcomes with underlying physical factors.

**Reproducibility:** All codes and scripts are publicly available at GitHub repository, enabling full reproducibility of the experiments and results.

## 3. Results

### 3.1 Model performance

The baseline Random Forest (RF) model achieved strong predictive performance, with a mean absolute error (MAE) of 0.0329 and a coefficient of determination ($R^2$) of 0.9272 on the test set. XGBoost (XGB) provided comparable accuracy, reaching MAE = 0.0366 and $R^2$ = 0.9165 as shown in **Table 1**. Cross-validation confirmed the robustness of both models, with only minor fluctuations between folds. RF displayed slightly superior accuracy on the test set, while XGB showed more stable generalization across folds. These findings confirm that ensemble-based methods are well-suited for capturing the nonlinear composition–property relationships underlying glass formation.

**Table 1. Performance comparison of RF and XGB models: Quantitative evaluation of Random Forest (RF) and XGBoost (XGB) models for predicting glass-forming ability (GFA). Metrics include test set mean absolute error (MAE), coefficient of determination ($R^2$), and cross-validation statistics. RF shows slightly better accuracy, while XGB demonstrates more stable generalization**

| Model | Test MAE | Test $R^2$ | CV MAE (± std) | CV $R^2$ (± std) |
|---|---|---|---|---|
| Random Forest | 0.0329 | 0.9272 | 0.0313 ± 0.0023 | 0.9300 ± 0.0127 |
| XGBoost | 0.0366 | 0.9165 | 0.0329 ± 0.0014 | 0.9309 ± 0.0096 |

## 3.2 Predicted vs. actual GFA

**Figure 1(a)** shows the scatter plot of predicted versus actual GFA values for the RF model. The strong clustering of data points along the diagonal line indicates excellent predictive agreement, with only minor deviations at extreme values. This result validates the ability of the model to reliably predict GFA across diverse ternary compositions.

### 3.3 Feature importance in Random Forest

Feature ranking from the RF model **Figure *1 (b)*** revealed fourteen dominant descriptors. The most important features were electronegativity variance, atomic size mismatch, and valence electron descriptors, which have direct physical relevance to glass formation. Secondary features, including cohesive energy and average ionic radius, also contributed meaningfully, indicating their influence on bonding strength and packing density. Together, these descriptors capture both chemical disorder and structural stability, explaining the strong predictive performance.

### 3.4 Cross-model comparison of features

**Figure *2*** compares the top 10 ranked features from RF and XGB. Both models consistently identified electronegativity variance, atomic size mismatch, and valence electron features as the most influential descriptors. Minor discrepancies appeared in the ranking of secondary descriptors such as cohesive energy and mean electronegativity, reflecting algorithm-specific sensitivities. This agreement across independent models underscores the robustness of the identified trends and strengthens confidence in their physical interpretation.

### 3.5 Extended feature ranking

An extended analysis of feature importance is provided in **Figure *3***, which compares the ranking of 14 descriptors across RF and XGB. While the relative weighting of secondary features differs slightly, the overall consistency of the dominant descriptors demonstrates that these chemical factors fundamentally govern GFA in ternary oxides.

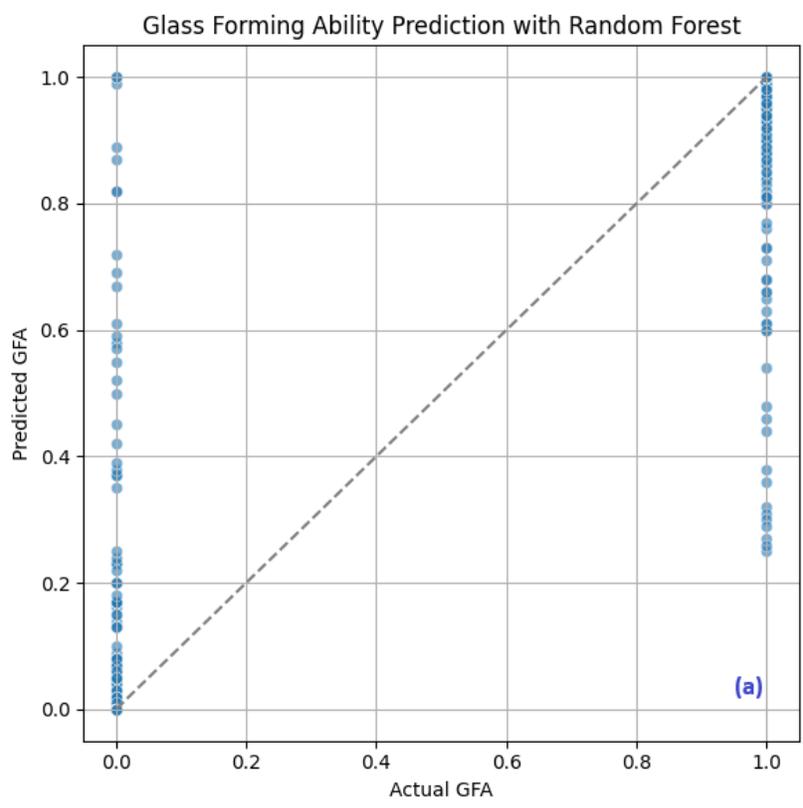

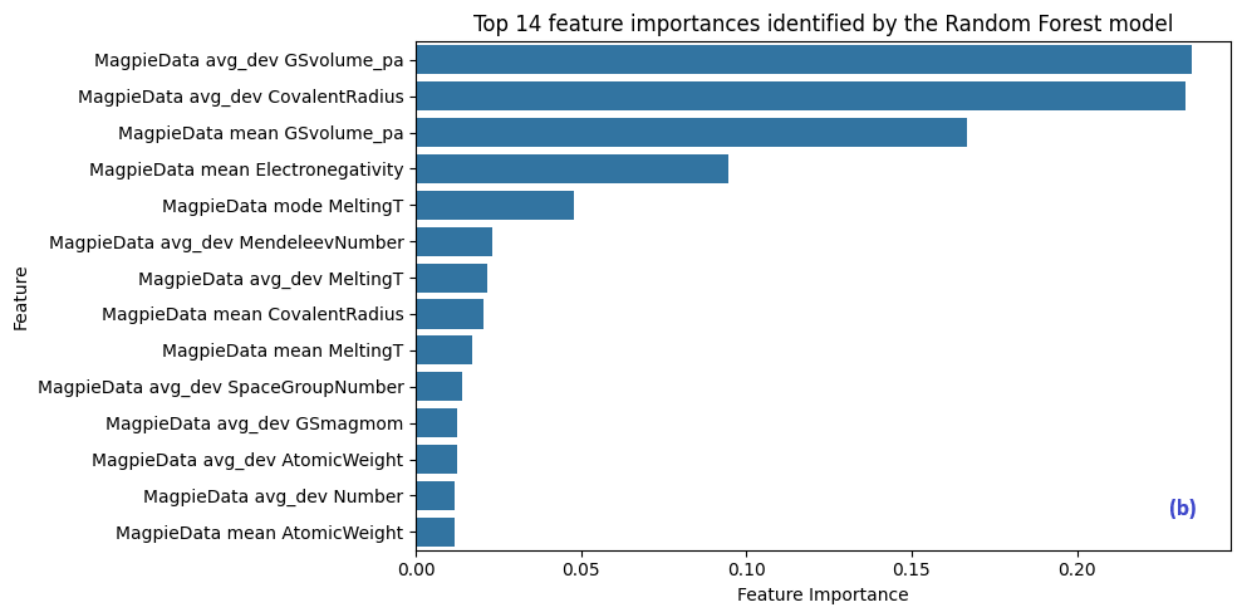

**Figure 1 (a) Predicted versus actual values of glass-forming ability (GFA) using the Random Forest model. The strong agreement along the diagonal line demonstrates the high predictive accuracy of the model. (b)Top 14 feature importances identified by the Random Forest model. Electronegativity variance, atomic size mismatch, and valence electron descriptors emerge as the most influential features governing GFA.**

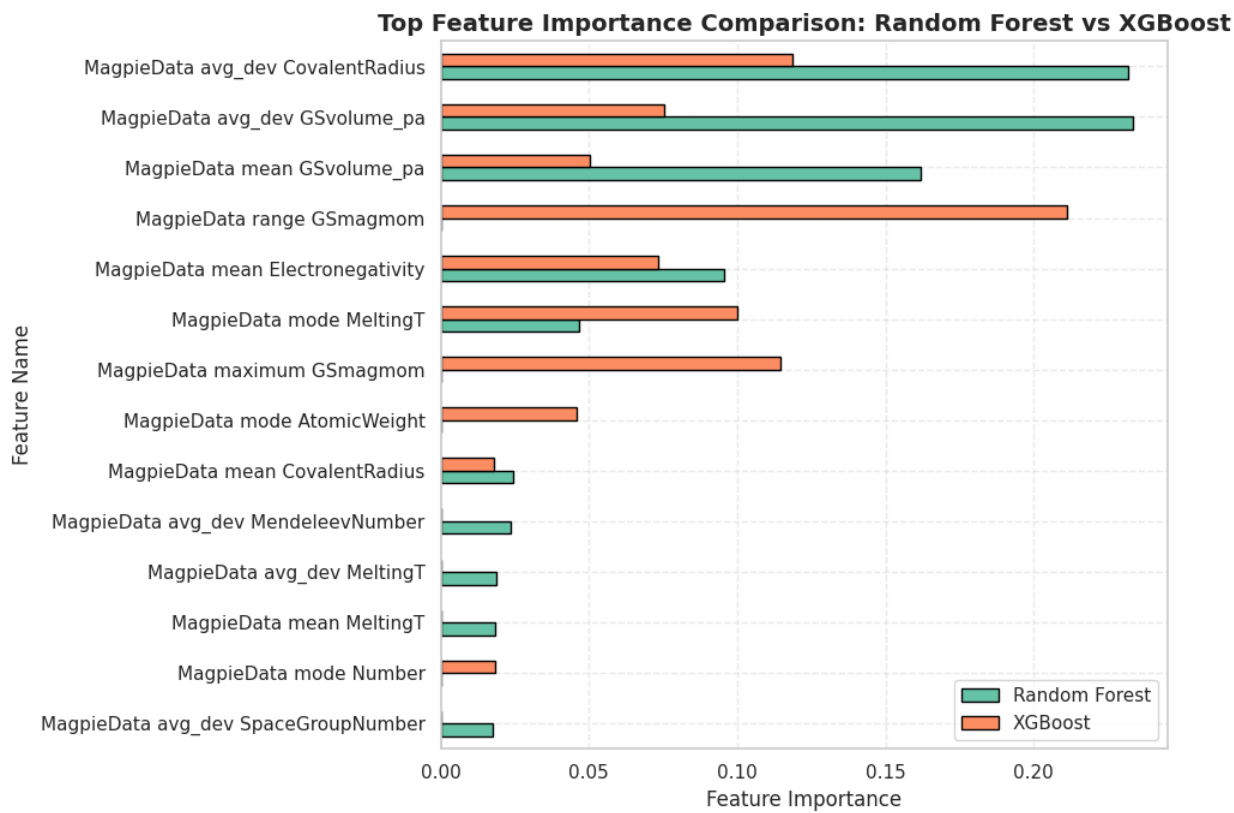

**Figure 2. Comparison of the top 10 features ranked by Random Forest (RF) and XGBoost (XGB). Both models consistently highlight electronegativity variance, atomic size mismatch, and valence electron–related descriptors as dominant factors controlling glass formation.**

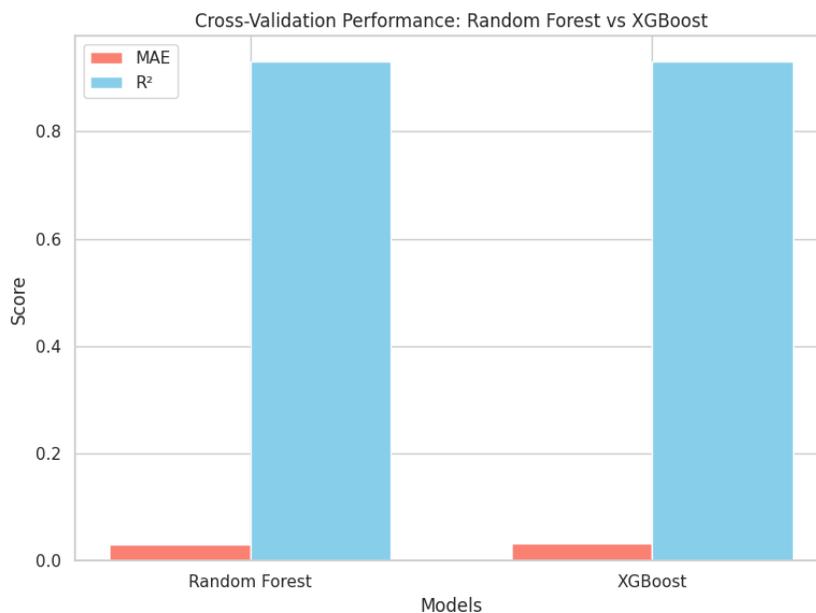

**Figure 3. Extended feature ranking analysis showing the relative importance of 14 descriptors in both RF and XGB. While the overall trends are consistent, subtle differences in feature prioritization emphasize the complementary strengths of the two models.**

## 4.Discussion

The results of this study demonstrate that ensemble ML models can predict the glass-forming ability (GFA) of ternary oxide glasses with high accuracy ($R^2 > 0.92$). More importantly, the feature importance analysis provides chemically interpretable insights that align well with established theories of glass formation. As shown in **Figure *1*(b)**, descriptors such as electronegativity variance and atomic size mismatch dominate the prediction, while **Figure *2*** highlights the consistency between Random Forest (RF) and XGBoost (XGB) in ranking the top 10 features. The extended feature ranking in **Figure *3*** further supports the robustness of these findings, showing that both models emphasize similar physical drivers of glass formation.

The physical interpretation of these features is summarized in **Table *2***. Electronegativity variance increases chemical disorder and reduces the likelihood of crystallization, which is consistent with classical concepts of glass stability proposed by Turnbull and Greer in metallic glass theory. Similarly, atomic size mismatch introduces packing frustration, a mechanism widely recognized in both metallic and oxide glasses. Valence electron descriptors, by influencing bond flexibility and network connectivity, echo established principles in glass science regarding the role of electronic structure in stabilizing amorphous states. The secondary contributions of cohesive energy, mean ionic radius, and average electronegativity complement these dominant descriptors, offering a more nuanced picture of how bonding strength and packing density interact in ternary systems.

Compared with prior ML studies, this work extends predictive modeling beyond metallic and binary systems into ternary oxide glasses. As highlighted in **Table *3***, previous research has achieved strong performance but with narrower compositional focus-metallic glasses [19, 20], binary alloys and oxides [8, 18] or Fe-based metallic glasses[21]. Our results show comparable or superior accuracy, while simultaneously providing chemically interpretable descriptors specific to oxides. This dual achievement of predictive power and interpretability constitutes the novelty of the present work.

Nonetheless, some limitations must be acknowledged. First, the dataset size remains modest and may not fully represent the vast chemical space of ternary oxides. Second, the descriptors are derived exclusively from compositional features (Magpie preset), without incorporating structural or thermodynamic information. While these descriptors are effective, their experimental measurability can be limited, and future studies should incorporate CALPHAD-based and structural descriptors for broader validation. Finally, the models have not yet been validated against newly synthesized experimental data, which is a necessary step for practical deployment.

Looking forward, several avenues for advancement are promising. Expanding datasets to cover broader compositional ranges will improve model generalizability. Incorporating thermodynamic and structural descriptors, such as formation enthalpy or network connectivity, will enrich the interpretability of results. Moreover, advanced architectures such as graph neural networks (GNNs) hold potential, as they can explicitly capture non-local structural information in glass networks—something beyond the reach of simple composition-derived features. Coupling ML predictions with experimental validation will ultimately accelerate the discovery of novel oxide glasses with tailored stability and functionality.

**Table 2. Physical interpretation of the most important features: Scientific meaning and role of the most influential compositional descriptors in glass formation. Features such as electronegativity variance and atomic size mismatch enhance chemical disorder and packing frustration, contributing to vitrification**

| Feature | Scientific meaning | Role in GFA |
|---|---|---|
| Electronegativity variance | Difference in electronegativity between cations | Enhances chemical disorder, suppresses crystallization |
| Atomic size mismatch | Disparity in ionic radii | Causes packing frustration, stabilizes the glassy state |
| Valence electron descriptors | Distribution of valence states | Controls bond flexibility and network stability |
| Mean cohesive energy | Average bonding strength | Reflects chemical stability, secondary influence |
| Average ionic radius | Mean cation size | Affects packing density and compactness |
| Electronegativity mean | Average electronegativity | Related to bond polarity, complements variance |

**Table 3. Comparison of this work with previous ML studies on GFA prediction Benchmarking this study against prior machine learning approaches for GFA prediction in metallic and oxide systems. Highlights the novelty of applying ensemble models to ternary oxide glasses with interpretable descriptors and high predictive accuracy**

| Study | System | Method | Accuracy ($R^2$) | Novelty / Limitation |
|---|---|---|---|---|
| Ghorbani et al. (2022)[19] | Metallic glasses | Thermodynamically guided RF | ~0.92 | Limited to BMGs |
| Liu et al. (2023)[20] | Metallic glasses | Regression + feature selection | >0.91 | Good interpretability, only BMGs |
| Pan et al. (2022, 2023)[8, 18] | Binary alloys/oxides | RF + ML optimization | ~0.90 | Narrow composition range |
| Bobadilla et al. (2025)[21] | Fe-based metallic glasses | ML + CALPHAD descriptors | ~0.69 | Limited accuracy, Fe-based only |
| **This work (2025)** | **Ternary oxide glasses** | **RF, XGB + feature analysis** | **>0.92** | **First systematic study on ternary oxides; identifies key descriptors** |

## 4. Conclusion

In this study, we demonstrated that ensemble machine learning methods-Random Forest (RF) and XGBoost (XGB)-can accurately predict the glass-forming ability (GFA) of ternary oxide glasses. Both models achieved excellent predictive accuracy ($R^2 > 0.92$, MAE $< 0.04$), confirming that composition-derived features are sufficient to capture the essential physical mechanisms governing vitrification.

Feature importance analysis consistently identified electronegativity variance, atomic size mismatch, and valence electron descriptors as the dominant factors influencing glass formation. Secondary contributions from cohesive energy, mean ionic radius, and mean electronegativity further highlighted the interplay between chemical bonding and structural packing. These findings not only validate the effectiveness of ML for predicting GFA but also provide chemically interpretable insights that align with established theories of glass science.

The novelty of this work lies in systematically extending ML-based predictive modeling to ternary oxide glasses-a class of materials less explored compared to metallic or binary systems-and in demonstrating that interpretable descriptors can be extracted directly from compositional data. This dual achievement of predictive accuracy and physical interpretability establishes ML as a reliable framework for guiding the design of novel oxide glasses.

Looking forward, expanding the dataset to include broader chemical spaces, incorporating thermodynamic and structural descriptors, and applying advanced architectures such as graph neural networks (GNNs) will further enhance predictive power. Coupling ML predictions with experimental validation will ultimately accelerate the discovery and optimization of new glassy materials with tailored properties for industrial applications.